\documentclass[preprint,review,12pt]{elsarticle}
 \usepackage{graphics}
 \usepackage{graphicx}

\usepackage{mathrsfs}
\usepackage{amssymb}

\usepackage{color}

\journal{Physics Letters B}
\begin{document}
\begin{frontmatter}



\title{Spin-Orbit Force from Lattice QCD}


\author[riken]{K.~Murano}
\author[ccs]{N.~Ishii}
\author[kyoto,ccs]{S.~Aoki}
\author[riken]{T.~Doi}
\author[riken,kipmu]{T.~Hatsuda}
\author[riken]{Y.~Ikeda}
\author[nihon]{T.~Inoue}
\author[ccs]{H.~Nemura}
\author[ccs]{K.~Sasaki}
\author[]{(HAL QCD Collaboration)}
\author[]{\includegraphics[width=.25\textwidth, bb=0 0 202 118]{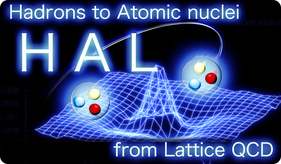}}

\address[riken]{Theoretical Research Division, Nishina Center, RIKEN,
  Saitama 351-0198, Japan}
\address[ccs]{Center for Computational Sciences, University of Tsukuba,
  Ibaraki 305-8571, Japan}
 \address[kyoto]{Yukawa Institute for Theoretical Physics, Kyoto University, Kitashirakawa Oiwakecho, Sakyo-ku, Kyoto 606-8502, Japan
 }
\address[kipmu]{Kavli IPMU, The University of Tokyo, Kashiwa 277-8583, Japan}
\address[nihon]{Nihon University, College of Bioresource Sciences,
  Kanagawa 252-0880, Japan}
\begin{abstract}
 We present a first attempt to determine nucleon-nucleon potentials in the
parity-odd sector, which appear in the $^1P_1$, $^3P_0$, $^3P_1$, $^3P_2$--$^3F_2$ channels, 
in $N_f=2$ lattice QCD simulations.
These potentials are constructed from the Nambu-Bethe-Salpeter
 wave functions  
 for $J^P=0^-,  1^-$ and  $2^-$, which correspond to the $A_1^-, \ T_1^-$ and $T_2^-\oplus E^-$ representation of the cubic group, respectively.
 We have found  a large and attractive spin-orbit potential $V_{\rm LS}(r)$ in the isospin-triplet
 channel,
 which is qualitatively consistent with the phenomenological determination from the
 experimental scattering phase shifts.   
 The  potentials obtained from lattice QCD 
 are used  to calculate  the scattering phase  shifts in the
 $^1P_1$, $^3P_0$,  $^3P_1$  and  $^3P_2$--$^3F_2$  channels.
 The strong attractive spin-orbit force and a weak repulsive central force in spin-triplet $P$-wave channels 
  lead to an attraction in the $^3P_2$ channel, which is related to the $P$-wave neutron paring in neutron stars.
\end{abstract}


\end{frontmatter}


\newcommand{\Sect}[1]{Section~\ref{#1}}
\newcommand{\Fig}[1]{Fig~\ref{#1}}
\newcommand{\Eq}[1]{Eq.~(\ref{#1})}
\newcommand{\agt}{\,\,\raisebox{.6ex}{$>$}\hspace{-0.8em}\raisebox{-.6ex}{$\sim$}\,\,}
\newcommand{\alt}{\,\,\raisebox{.6ex}{$<$}\hspace{-0.8em}\raisebox{-.6ex}{$\sim$}\,\,}
\newpage
\section{Introduction}\label{intro}

 A study of the nuclear force in QCD is a first step toward the understanding of hadronic properties beyond single hadrons.
 In  addition,  the nuclear  force  plays  a  key role  in  describing
 properties     of     atomic     nuclei     and     neutron     stars
 \cite{Epelbaum:2008ga,Machleidt:2011zz}.
 In lattice QCD, the standard method to study hadronic interactions is
 the finite volume method \cite{Luscher:1990ux} by calculating the
 scattering phase shift\cite{Fukugita:1994na,
   Beane:2008dv,Beane:2010em,Yamazaki:2009ua,Yamazaki:2012hi,Beane:2013br}.
   Recently, a new 
 method to extract hadronic interactions
 from (lattice) QCD has been proposed, where non-local potential can
 be defined through the Nambu-Bethe-Salpeter (NBS) wave function.
 The method has been successfully applied to the nuclear forces
 \cite{Ishii:2006ec,Aoki:2009ji,HALQCD:2012aa}.
 It has been extended to various other systems such as hyperon-nucleon
 (YN), hyperon-hyperon (YY),
meson-baryon, 
and the three-nucleons\cite{Aoki:2012tk}.
In Ref.~\cite{Kurth:2013tua}, an explicit comparison between the
finite volume method and the potential method is made 
in the case of $\pi\pi$ scattering phase shifts,
where a good agreement is obtained.

In the case of the nucleon-nucleon (NN) system, for example, non-local
potentials are defined through the Schr\"odinger equation for the NBS
wave function.  Below the pion production threshold, the non-local NN
potential can be expanded by the number of derivatives with respect to
its non-locality of the relative coordinate. The leading order (LO)
terms are the spin-independent central potential $V_0(r)$, the
spin-dependent central potential $V_{\sigma}(r)$ and the tensor
potential $V_{\rm T}(r)$, while the next-to-leading order (NLO) term
is the spin-orbit potential $V_{\rm LS}(r)$.  Up to the NLO, there are
altogether 8 independent local potentials, $V^{I=0,1}_{0,\sigma,{\rm
    T},{\rm LS}}$ where $I$ denotes the total
isospin\cite{Aoki:2009ji}.  In the previous
studies\cite{Ishii:2006ec,Aoki:2009ji,HALQCD:2012aa},
  $V_{\rm
  C;S=0}^{I=1} \equiv V_0^{I=1} - 3 V_\sigma^{I=1}$, $V_{\rm
  C;S=1}^{I=0} \equiv V_0^{I=0} + V_{\sigma}^{I=0}$ and $V_T^{I=0}$
have been determined from the NBS wave functions in $S$ and $D$ waves
(the parity-even sector) at various lattice parameters.
For the complete determination, however, we must employ the NBS wave functions in $P$ and $F$ waves (the parity-odd sector) with non-zero relative momentum of the NN system, 
where  relevant channels are $^1P_1$, $^3P_0$, $^3P_1$ and $^3P_2$--${^3F_2}$.
 The corresponding potentials are given by
\begin{eqnarray}
  V(r;\ ^1P_1) &=& V^{I=0}_{{\rm C},S=0}(r) \equiv V^{I=0}_0(r) - 3V^{I=0}_{\sigma}(r) 
    \label{eq:effective-potentials-1}\\
 V(r;\ ^3P_0) &=& V^{I=1}_{{\rm C},S=1}(r) - 4V^{I=1}_{\rm T}(r) - 2 V^{I=1}_{\rm LS}(r)
    \label{eq:effective-potentials-2}
  \\
  V(r;\ ^3P_1) &=& V^{I=1}_{{\rm C},S=1}(r)  + 2V^{I=1}_{\rm T}(r) - V^{I=1}_{\rm LS}(r)
   \label{eq:effective-potentials-3}\\
 V(r;\ ^3(P,F)_2) &=& \left(
\begin{array}{ll}
V(r;\ ^3P_2) & \frac{6\sqrt{6}}{5}V^{I=1}_{\rm T}(r)\\
\frac{6\sqrt{6}}{5}V^{I=1}_{\rm T}(r) & V(r;\ ^3F_2) \\
\end{array} 
 \right)
 \nonumber 
\end{eqnarray}
 with $V^{I=1}_{{\rm C},S=1}(r) =V^{I=1}_0(r) + V^{I=1}_{\sigma}(r)$,
 $V(r;\ ^3P_2) =V^{I=1}_{{\rm C},S=1}(r)  - \frac{2}{5}V^{I=1}_{\rm T}(r) + V^{I=1}_{\rm LS}(r)$
 and
$V(r;\ ^3F_2)=V^{I=1}_{{\rm C},S=1}(r) - \frac{8}{5} V^{I=1}_{\rm T}(r) - 4 V^{I=1}_{\rm LS}(r)$.
 The  LS force has  close relation to the
  spin-orbit splittings of the nuclear  spectra and the nuclear magic  numbers \cite{Bohr}.
 Large neutron-neutron attraction due to the LS force in the $^3P_2$--${^3F_2}$
 channel leads to the $P$-wave superfluidity in the stellar environment such as the
  neutron star interiors \cite{tamagaki,hoffberg,Baldo:1998ca}. 
  It also  affects the cooling properties of neutron stars \cite{Page:2013hxa}.
 
The present paper is organized  as follows.
In \Sect{sect.definition},  we give  a brief review  of our  method to
obtain NN  potentials on the lattice.
\Sect{sect.nbs.wave}  is  devoted  to the discussion on  the   NBS  wave
functions with non-zero angular momenta.  In  \Sect{sect.numeric}, we  present numerical  results of the 
 potentials. The scattering phases and mixing parameters
in the $^1P_1$, $^3P_0$, $^3P_1$, and $^3P_2$--$^3F_2$ channels calculated form these potentials are also presented.

\section{NN potential in QCD}
\label{sect.definition}

Below the  inelastic threshold of the NN system ($W <  2m_N + m_{\pi}$),  
we can define the non-local but energy-independent potential as\cite{Aoki:2009ji}
\begin{equation}
  \left(
    k^2/m_N - H_0
  \right)
  \phi(\vec{r};W)
  =
  \int d^3 r^\prime \, U(\vec{r},\vec{r^\prime}) \phi(\vec{r^\prime},W),
  \label{eq:schrodinger-like}
\end{equation}
from the Nambu-Bethe-Salpeter (NBS)  wave function in  the center of  mass (CM)
frame defined by
\begin{eqnarray}
	\phi_{\alpha\beta}(\vec{r};W)
        \equiv
        \langle 0 \vert
        p_\alpha (\vec{x})
        n_\beta (\vec{y})
        \vert B=2, W \rangle, \quad (\vec{r}=\vec{x}-\vec{y}),
        \label{eq:NBS}
\end{eqnarray}
where  $H_0 \equiv  -\nabla^2/m_N$ with the nucleon mass $m_N$,
$p_\alpha(\vec x)$,  $n_\beta(\vec y)$  denote  local composite
nucleon  operators with  spinor  indices $\alpha$, $\beta$  in  Dirac
representation restricted to $\alpha,\beta=0,1$, $\vert B=2, W \rangle $ is 
the state with baryon-number $B=2$ and vanishing total momentum with the total energy $W$. 
 Due to the confinement of quarks and gluons,  $\phi(\vec{r};W)$ for large $|\vec{r}|$
 reduces to the relative wave function of the non-interacting nucleons, so that
 $W$ can be written as $W=2\sqrt{k^2+m_N^2}$ 
 with $k \equiv \vert \vec k\vert$ being the asymptotic relative momentum 
  between the nucleons.
The identity 
$  U(\vec   r,\vec {r^\prime})  =  V(\vec r,\vec  \nabla)  \delta(\vec  r -  \vec{r^\prime}) $ 
leads to the derivative expansion up to the NLO order:
\begin{equation}
  V^I(\vec r,\vec \nabla)
  =
  V^I_{0}(r)
  + V^I_{\sigma}(r) \vec \sigma_1\cdot\vec\sigma_2
  + V^I_{\rm T}(r) S_{12}
  + V^I_{\rm LS}(r) \vec L\cdot\vec S
  + O(\nabla^2),
 \label{eq:OM-decomposition} 
\end{equation}
where
 $S_{12}\equiv  3(\vec  \sigma_1\cdot\vec r)(\vec  \sigma_2\cdot
\vec r)/r^2  - \vec  \sigma_1\cdot\vec\sigma_2$, $\vec S  \equiv (\vec
\sigma_1 + \vec  \sigma_2)/2$ and $\vec L \equiv i  \vec r \times \vec
\nabla$ denote the  tensor, the total spin and  the (relative) orbital
angular momentum operators, respectively, and $I=0,1$  is the total isospin of the two nucleons.

 These NN potentials can be extracted by solving
 \Eq{eq:schrodinger-like}   with \Eq{eq:OM-decomposition} for NBS wave functions projected to appropriate quantum numbers.  

The NBS wave function in eq.~(\ref{eq:NBS})  and thus the non-local potential defined in eq.~(\ref{eq:schrodinger-like}) depend on the choice of nucleon operators $p(x)$ and $n(x)$.
   This is not surprising since the potential is not a physical observable and depends on how it is defined.  On the other hand, it has been shown that the NBS wave function carries  information of the scattering phase shift in its asymptotic behavior at large $r$ \cite{Lin:2001ek, Aoki:2005uf,Ishizuka2009a,Aoki:2009ji}. This property holds for an arbitrary choice of operators to define the NBS wave function, as long as basic properties in quantum field theories such as locality and unitarity are satisfied. Therefore, by construction, non-local potentials in different definitions give same and  correct phase shifts through the Schr\"odinger equation in eq.~(\ref{eq:schrodinger-like}). 
Despite that the above points have been already stated clearly and explicitly in our previous papers (see, e.g. \cite{Aoki:2009ji}), similar remarks appear repeatedly in later literature (see e.g. \cite{Birse:2012ph}).

   In our study, we take local interpolating  operators for the nucleon as 
$
  p(x)
  \equiv
  \epsilon_{abc}
  \left(u_a^T(x)C\gamma_5d_b(x)\right)
  u_c(x)$,
and
$n(x)
  \equiv
  \epsilon_{abc}
  \left(u_a^T(x)C\gamma_5d_b(x)\right)d_c(x)$,
to define the NBS wave function, where $a,b,c$ denotes the  color indices.
 In our actual calculation, the derivative expansion of the non-local potential 
 in eq.~(\ref{eq:OM-decomposition}) is truncated  at fixed order (the NLO in the present paper); 
 then  the resultant phase shifts  are valid only in a certain energy interval
 which depends on the order of the truncation and the choice of the interpolating operator.  
The former dependence  has been investigated for the NN case\cite{Murano:2011nz} and the $\pi\pi$ case\cite{Kurth:2013tua}, while the latter dependence is still left for future studies. 

\section{NBS wave functions with non-zero angular momenta}
\label{sect.nbs.wave}
The  NBS  wave function  for the ground  state can be
extracted from  nucleon four-point functions at  large $t$ as
\begin{eqnarray}
  G_{\alpha\beta}(\vec{r},t-t_0; \mathscr{J})
  &\equiv&
  \frac{1}{V} \sum_{\vec{y}}
  \left\langle 0 \left|
      T\left[
        p_{\alpha}(\vec{r}+\vec{y},t)
        n_{\beta}(\vec{y},t) \mathscr{J}(J^{P},S;t_0)
      \right]
    \right| 0 \right\rangle,
  \nonumber
  \\
  &=&
  \sum_{m}
  \phi(\vec{r};W_m)
  \left\langle m\left|
      \mathscr{J}(J^{P},S;0)
    \right| 0 \right\rangle
  e^{-W_m(t-t_0)},\label{eq:four-point}
\end{eqnarray}
where  two-nucleon  source  $\mathscr{J}(J^{P};t_0)$  located at  the  time-slice
$t=t_0$ is used to control the quantum numbers such as $J^P$, and 
$W_m$ denotes the  energy of the intermediate state  $\vert m\rangle$.
The summation over $\vec{y}$ is performed to select states
with the vanishing total spatial momentum.

To construct sources coupled to the parity-odd states, we employ
wall sources with non-zero momentum given by
\begin{equation}
  \mathcal{J}_{\alpha\beta}(f_{i})
  \equiv
  \sum_{\vec x_1,\cdots,\vec x_6}
  \bar{P}_{\alpha}(\vec x_1,\vec x_2,\vec x_3)
  \bar{N}_{\beta }(\vec x_4,\vec x_5,\vec x_6)
  f_{i}(\vec x_3 - \vec x_6),
\end{equation}
with
$
  \bar{P}_{\alpha}(\vec x_1,\vec x_2,\vec x_3)
  \equiv
  \epsilon_{abc}
  \left(
    \bar u_a(\vec x_1)
    C\gamma_5
    \bar d_b^T(\vec x_2)
  \right)
  \bar u_{c,\alpha}(\vec x_3)
$,
and
$
  \bar{N}_{\alpha}(\vec x_4,\vec x_5,\vec x_6)
$
$
  \equiv
$
$
  \epsilon_{abc}
  \left(
    \bar u_a(\vec x_4)
    C\gamma_5
    \bar d_b^T(\vec x_5)
  \right)
  \bar d_{c,\alpha}(\vec x_6)
$, 
where
$f_{i}(\vec r)$ denotes a plane wave  with the spatial momentum
parallel or anti-parallel to a coordinate axis as
\[
  \begin{array}{lll}
    f_0(\vec r) \equiv \exp(+ i2\pi x/L),
    &
    f_1(\vec r) \equiv \exp(+ i2\pi y/L),
    &
    f_2(\vec r) \equiv \exp(+ i2\pi z/L),
    \\
    f_3(\vec r) \equiv \exp(- i2\pi x/L),
    &
    f_4(\vec r) \equiv \exp(- i2\pi y/L),
    &
    f_5(\vec r) \equiv \exp(- i2\pi z/L).
  \end{array}
\]

An  element  $g$  of  cubic  group $\mathcal{O}$  acts  on  these  six
functions as permutation,
$  f_{i}(g \vec r)
  =
  U_{ij}(g)
  f_{j}(\vec r),
$
where $U(g)$ is the representation  matrix of the cubic group whose explicit form can be
generated by the basic matrices,
\begin{equation}
  U(C_{4y})
  \equiv
  \left[
    \begin{array}{cccccc}
      0 & 0 & 1 & 0 & 0 & 0 \\
      0 & 1 & 0 & 0 & 0 & 0 \\
      0 & 0 & 0 & 1 & 0 & 0 \\
      0 & 0 & 0 & 0 & 0 & 1 \\
      0 & 0 & 0 & 0 & 1 & 0 \\
      1 & 0 & 0 & 0 & 0 & 0
    \end{array}
  \right],\,
  U(C_{4z})
  \equiv
  \left[
    \begin{array}{cccccc}
      0 & 0 & 0 & 0 & 1 & 0 \\
      1 & 0 & 0 & 0 & 0 & 0 \\
      0 & 0 & 1 & 0 & 0 & 0 \\
      0 & 1 & 0 & 0 & 0 & 0 \\
      0 & 0 & 0 & 1 & 0 & 0 \\
      0 & 0 & 0 & 0 & 0 & 1
    \end{array}
  \right],
\end{equation}
with $C_{4y}$  and $C_{4z}$ being  the rotations by +90  degrees around
y-axis and z-axis, respectively, and  $U(g)$ for other $g \in \mathcal{O}$
are obtained  by multiplying these  two matrices in  suitable orders.
For instance,  the representation matrix  for $C_{4x}$, a  rotation by
+90 degrees  around the x-axis,  is obtained as $U(C_{4x})  = U(C_{4y})
U(C_{4z}) U^{-1}(C_{4y})$.
The spatial reflection $R$ corresponds to 
$f_0 \leftrightarrow f_3$, $f_1 \leftrightarrow f_4$,
and $f_2 \leftrightarrow f_5$, which leads to $6 \times 6$ representation matrix
$U(R)=\left(
  \begin{array}{cc}
    0_{3\times 3}   & I_{3\times 3}   \\
    I _{3\times 3}  & 0_{3\times 3}   \\
  \end{array}
\right)
$. 
Analysis based on the group characters shows that $U(g)$ is reduced to
the  direct sum  of  irreducible representations  $T_1^- \oplus  A_1^+
\oplus E^+$.

Together with the transformation property of the quark field $\bar q(\vec x) \mapsto
\bar  q(g^{-1}\vec  x)  S(g^{-1})$  where $S(g)$  denotes the standard  rotation matrix acting on
upper   two  components  in   the  Dirac representation,
we have
\begin{eqnarray}
  \mathcal{J}_{\alpha\beta}(f_{i})
  &\mapsto&
  \sum_{\vec x_1,\cdots,\vec x_6}
  \bar P_{\alpha'}(g^{-1}\vec x_1,\cdots,g^{-1}\vec x_3)
  \bar N_{\beta'} (g^{-1}\vec x_4,\cdots,g^{-1}\vec x_6)
  f_{i}(\vec x_3 - \vec x_6)
  \nonumber
  \\
  &&
  \times
  S_{\alpha'\alpha}(g^{-1})
  S_{\beta'\beta}  (g^{-1})
  =
  U_{ij}(g)
  \mathcal{J}_{\alpha'\beta'}(f_{j})
  S_{\alpha'\alpha}(g^{-1})
  S_{\beta'\beta}  (g^{-1}).
\end{eqnarray}
Therefore  our   momentum      wall  source
$\mathcal{J}_{\alpha\beta}(f_i)$  covers  $ T_1^-\oplus  A_1^+
\oplus E^+ $ ($\simeq 1^- \oplus 0^+ \oplus 2^+$) for the spin singlet
sector and  $(T_1^-\oplus A_1^+ \oplus E^+)\otimes T_1  = A_1^- \oplus
T_1^-  \oplus (T_2^-  \oplus  E^-) \oplus  T_1^+  \oplus T_1^+  \oplus
T_2^+$ ($\simeq 0^- \oplus 1^- \oplus 2^- \oplus 1^+ \oplus 1^+ \oplus
2^+$)  for  the  spin  triplet  sector.  \footnote{See  Appendix  A  in
  Ref.\cite{Murano:2011nz}  for a  decomposition of  a product  of two
  irreducible representations of the  cubic group and for the relation
  of irreducible representations between the cubic group and $SO(3)$.}

We introduce several projections to  fix quantum numbers of the  source    operator
$\mathcal{J}_{\alpha\beta}(f_{i})$.
The projection for the total  angular
momentum $J$ is given by
\begin{eqnarray}
  \mathcal{P}^{(J)}[\mathcal{J}_{\alpha\beta}(f_{i})]
  &\equiv&
  \frac{d^{(J)}}{24}
  \sum_{g\in\mathcal{O}}
  \chi^{(J)}(g^{-1})
  \cdot
  U_{ij}(g)
  \mathcal{J}_{\alpha'\beta'}(f_{j})
  S_{\alpha'\alpha}(g^{-1})
  S_{\beta'\beta}  (g^{-1})
  \nonumber
  \\
  &=&
  P^{(J)}_{\alpha\beta i;\alpha'\beta' j}
  \cdot
  \mathcal{J}_{\alpha'\beta'}(f_{j}),
\end{eqnarray}
where  $d^{(J)}$  and $\chi^{(J)}(g)$  denote  the  dimension and  the
character of  the irreducible representation  $J$ of the  cubic group,
and 
the  $24\times 24$ matrix  $P^{(J)}_{\alpha\beta i;  \alpha'\beta' j}$
is the projection matrix onto the irreducible representation $J$.
Similar  considerations give the projection matrices  for the
parity
\begin{equation}
  P^{(P=\pm)}_{\alpha\beta i;\alpha'\beta' j}
  \equiv
  \frac1{2}
  \left(
    \delta_{ij}
    \pm
    U_{ij}(R)
  \right)
  \cdot
  \delta_{\alpha\alpha'}
  \delta_{\beta\beta'}.
\end{equation}
as well as the total spin $S$
\begin{eqnarray}
  P^{(S=0)}_{\alpha\beta i;\alpha'\beta' j}
  &\equiv&
  \frac1{4}
  \left(
    1 - \vec\sigma_1\cdot\vec\sigma_2
  \right)_{\alpha\beta;\alpha'\beta'}
  \cdot
  \delta_{ij},
 \\\nonumber
  P^{(S=1)}_{\alpha\beta i;\alpha'\beta' j}
  &\equiv&
  \frac1{4}
  \left(
    3 + \vec\sigma_1\cdot\vec\sigma_2
  \right)_{\alpha\beta;\alpha'\beta'}
  \cdot
  \delta_{ij}.
\end{eqnarray}
The  projection matrices  for $J_z$  are  defined based  on the  $C_4$
subgroup  of   the  cubic  group   which  consists  of   multiples  of
$C_{4z}$ as
\begin{equation}
  P^{(J_z=M)}_{\alpha\beta i;\alpha'\beta' j}
  \equiv
  \frac1{4}
  \sum_{n=0,1,2,3}
  e^{i(\pi/2)M n}
  \cdot
  U_{ij}((C_{4z})^n)
  S_{\alpha\alpha'}((C_{4z})^{-n})
  S_{\beta\beta'}((C_{4z})^{-n}).
\end{equation}

Since a product  of these projection matrices \footnote{The result
  does  not  depend  on  the  order of  multiplications,  since  these
  projection matrices are mutually commutative with each other.} 
\begin{equation}
  P^{(J,J_z,P,S)}_{\alpha\beta i;\alpha'\beta' j}
  \equiv
  \left(P^{(J)} P^{(J_z)} P^{(P)} P^{(S)} \right)_{\alpha\beta i;\alpha'\beta' j}.
\end{equation}
has the   property   $(P^{(J,J_z,P,S)})^2=P^{(J,J_z,P,S)}$,  the
eigenvalues of $P^{(J,J_z,P,S)}$ are either 0 or 1.
We   diagonalize   $P^{(J,J_z,P,S)}$   to  obtain   its   eigenvectors
$\eta^{(J,J_z,P,S); n}_{\alpha\beta i}$ with eigenvalue 1 as
$  P^{(J,J_z,P,S)}_{\alpha\beta i; \alpha'\beta' j}
  \eta^{(J,J_z,P,S); n}_{\alpha'\beta' j}
  =
  \eta^{(J,J_z,P,S); n}_{\alpha\beta i},
$
which is used to perform the projection of our two-nucleon source as
\begin{equation}
 \mathscr{J}(J^{P},J_z,S)
  \equiv
  \mathcal{J}_{\alpha\beta}(f_i)
  \eta^{(J,J_z,P,S);n}_{\alpha\beta i},
\end{equation}
where $n$ is  used to describe a possible degeneracy  of a given set of
quantum numbers $J,J_z,P,S$.
By    construction,     a   state
$\mathcal{J}(J^P,J_z,S)|0\rangle$ has  conserved quantum numbers $J^P,
J_z$ and $S$ in the two-nucleon sector.

\section{Numerical Results}
\label{sect.numeric}
\subsection{Lattice Setup}
In this study, we employ $N_f=2$   full  QCD configurations generated  by the CP-PACS  Collaboration on a  $16^3 \times 32$  lattice  with the RG improved  gauge action  (Iwasaki  action)  
at  $\beta =1.95$ and 
 with  the $\mathcal{O}(a)$-improved   Wilson  quark
(clover)  action  at $\kappa= 0.1375$ and $C_{\rm  SW}=1.53$,
which gives the  lattice spacing $a= 0.1555(17)$ fm,  the spatial extension $L = 16a = 2.489(27)$
fm,  the pion mass $m_{\pi} \simeq 1133$ MeV  and the nucleon mass $m_N  \simeq 2158$ MeV\cite{hep-lat/0105015}.  
The Dirichlet boundary condition along  the temporal direction is employed
to  generate quark  propagators to  avoid contamination  from backward
propagation of nucleons with negative parity.

With the projection defined in the previous section, 
we obtain the NBS  wave functions for $T_1^-$ in
the  spin singlet  sector and   for  $A_1^-, \ T_1^-, \ (E^-\oplus T_2^-)$ in the spin triplet sector.
In order  to improve statistics,  we perform the  measurement on $32$
source points by temporally shifting the location of the source,
in addition to averages over the charge-conjugation/time-reversal
transformations. We further reduce noises by  the average over the cubic group, as will be discussed later. 

To  construct  the  potentials,   we  use  the  time-dependent  method
\cite{HALQCD:2012aa}  with  a slight  modification  to  cope with  the
deviation of relativistic dispersion relation due to heavy quark mass,
as explained in the following subsection.
The  nearest neighbor  derivative is  used to  define  the discretized
Laplacian,  while the  symmetric derivative  is employed  to  define the
operator $\vec L$.
  To define $S_{12}$ and $\vec L$ on the periodic lattice, we take the
  origin of $\vec r$ to be the nearest periodic copy of the origin.
  On the spatial boundaries, i.e., $x = \pm L/2$ or $y =
  \pm L/2$ or $z = \pm L/2$, however,
  these operators  are still  ill-defined.  We therefore  exclude data on these boundaries
in our analysis. 
To  extract potentials,  we employ $t-t_0=8$, which is determined from $t$ dependencies of  potentials and phase shifts.

\subsection{Modified time-dependent method}
In Ref.~\cite{HALQCD:2012aa}, 
the time-dependent method has been proposed to  extract the potential directly from the four-point functions.   The  method indeed gives more  accurate and stable results,  since  it does not rely on the ground-state saturation at large $t$.
We therefore employ this method also in this paper.

For a coarse lattice, however, the heavy quark may violate the relativistic dispersion relation of a single nucleon
 as $E^2 \simeq m_N^2  + \alpha \vec k^2$ with  $\alpha \neq 1$ \cite{ElKhadra:1996mp}.
In such a case, 
the formula in Ref.~\cite{HALQCD:2012aa} receives a slight modification as
\begin{equation}
  \left\{
    \frac{1}{\alpha}\left(
      \frac{1}{4m_N}\frac{\partial^2}{\partial t^2} -\frac{\partial}{\partial t}
    \right) - H_0
  \right\} {\cal R}(\vec r, t;{\mathscr J})
  =
  \int d^3 r'\ 
  U(\vec r,\vec r')
  {\cal R}(\vec r', t;{\mathscr J}),
  \label{eq:t-dep}
\end{equation}
where   ${\cal R}(\vec    r,   t;{\mathscr   J})   \equiv    G(\vec   r,   t;
\mathscr{J})/(e^{-m_N  t})^2$.


In our simulation,
the dispersion relation for the  nucleon can be fitted well with $\alpha   =  0.88(1)$  ( $\chi^2/{\rm  d.o.f}   =  2.6$ ) at  $m_N=2152(3)$  MeV,  showing no sign   of  higher  order
contributions  in $k^2$  for $k^2 \le 1.25 {\rm [GeV^2]}$ ($ka \le \sqrt{5} \times 2\pi/L$) within statistical  errors.

\subsection{Extractions of potentials}
\begin{figure}[tb]
  \includegraphics[width=0.5\textwidth]{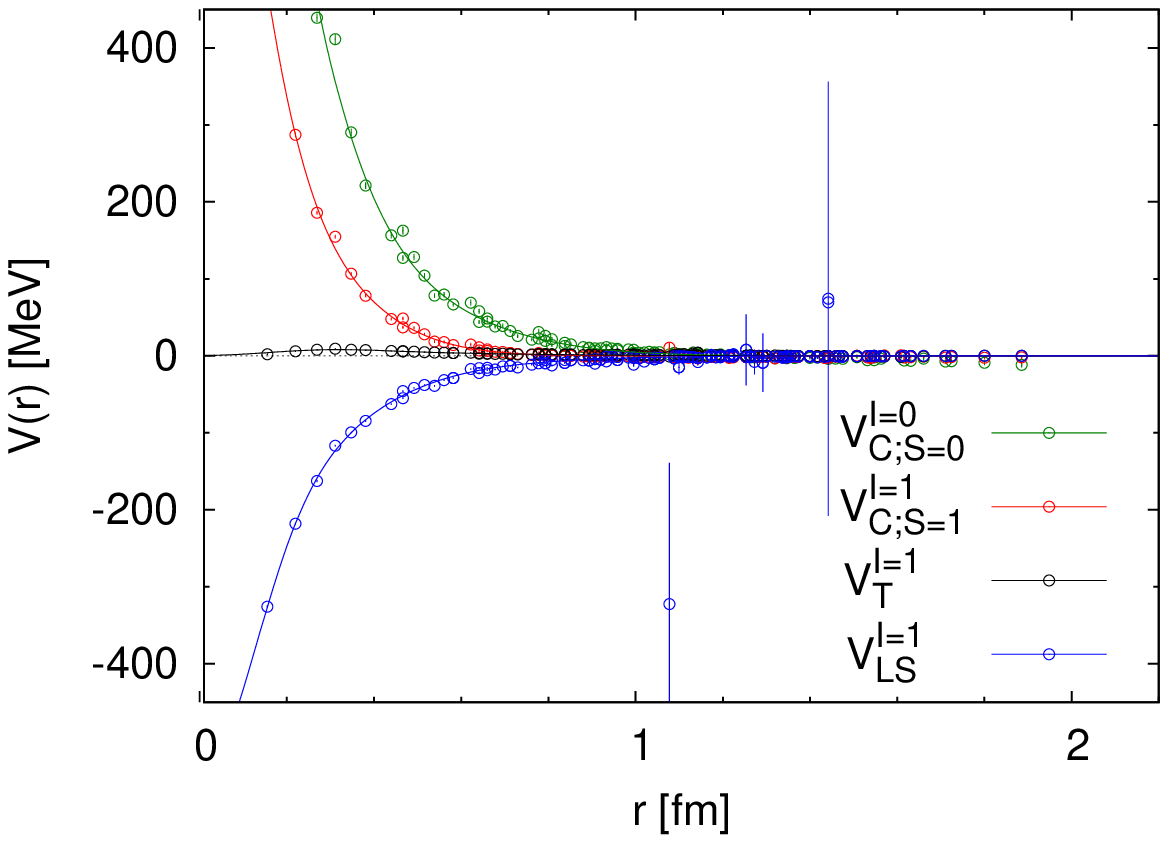}
  \includegraphics[width=0.5\textwidth]{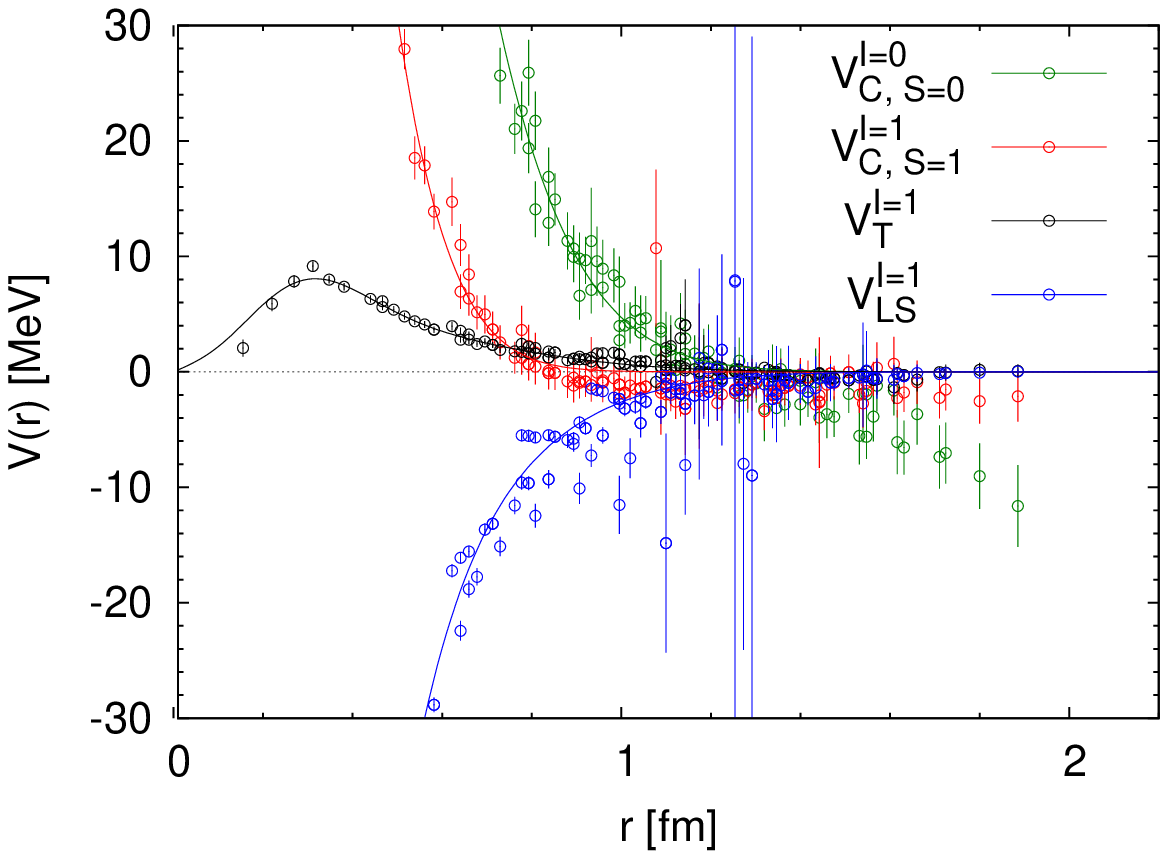}
  \caption{Central ($S=0$  and $1$), tensor  and spin-orbit potentials
    in parity-odd sector   obtained by lattice  QCD (left),  and their enlargements  (right).
}
 \label{fig:VcVTVLS}
\end{figure}

The potential for the spin-singlet sector at NLO can be easily extracted from the equation
\begin{equation}
V_{{\rm C},S=0}^{I=0}(r) \langle {\cal R}(\vec r,t;{\mathscr J}), {\cal R}(\vec r,t;{\mathscr J}) \rangle = \langle {\cal R}(\vec r,t;{\mathscr J}), (D_t - H_0)  {\cal R}(\vec r,t;{\mathscr J}) \rangle 
\end{equation} 
for ${\mathscr J} = {\mathscr J}(T_1^-)$, dominated by  $^1P_1$,
where $\alpha D_t = \frac{1}{4m_N}\frac{\partial^2}{\partial t^2}-\frac{\partial}{\partial t}$, and
we define an inner product with an average over the cubic group as $\langle F(\vec r), H(\vec r)\rangle \equiv \sum_{g\in {\cal O}} F_{\beta\alpha}^*(g\,\vec r) H_{\alpha\beta}(g\, \vec r)$,
which reduces statistical noises of potentials.  Note that  here and in the following we use the fact that local potentials, $V^{I}_{{\rm C},S}$, $V^I_{\rm T}$ and  $V^I_{\rm LS}$, are invariant under the rotation $g$ in the cubic group.
The result for $V_{C,S=0}^{I=0}(r) $ is plotted  in Fig.\ref{fig:VcVTVLS} by green circles, which
shows a strong repulsion at short distances.

For  the  spin-triplet  sector,  three unknown functions up to NLO can be determined from the equation
\begin{eqnarray}
V_{{\rm C};S=1}^{I=1}(r)  F_{\rm C}^{\mathscr J}(r) + V_{\rm T}^{I=1}(r) F_{\rm T}^{\mathscr J}(r) + V_{\rm LS}^{I=1}(r) F_{\rm LS}^{\mathscr J}(r)
&=& K^{\mathscr J}(r)
\label{eq:triplet}
\end{eqnarray}
for three different sources, ${\mathscr J} ={\mathscr J}(A_1^-)$, ${\mathscr J}(T_1^-)$, ${\mathscr J}(E^-)$( or ${\mathscr J}(T_2^- )$ ), dominated by $^3P_0$,  $^3P_1$ and $^3P_2$--$^3F_2$, respectively, 
where
\begin{eqnarray}
F_{\rm C}^{\mathscr J}(r) &\equiv& \langle {\cal R}(\vec r,t;{\mathscr J}), {\cal R}(\vec r,t;{\mathscr J})\rangle ,\ 
F_{\rm T}^{\mathscr J}(r) \equiv \langle {\cal R}(\vec r,t;{\mathscr J}), S_{12}\, {\cal R}(\vec r,t;{\mathscr J})\rangle , \nonumber \\
F_{\rm LS}^{\mathscr J}(r) &\equiv&  \langle {\cal R}(\vec r,t;{\mathscr J}), \vec L\cdot \vec S \, {\cal R}(\vec r,t;{\mathscr J})\rangle , 
\nonumber \\
K^{\mathscr J}(r) &\equiv& \langle {\cal R}(\vec r,t;{\mathscr J}), (D_t - H_0)  {\cal R}(\vec r,t;{\mathscr J}) \rangle .
\nonumber
\end{eqnarray}
In   Fig.~\ref{fig:VcVTVLS},   we also plot  $V^{I=1}_{{\rm C};S=1}(r)$(red), 
$V^{I=1}_{\rm  T}(r)$(black) and $V^{I=1}_{\rm  LS}(r)$(blue), obtained from  $A_1^-, \ T_1^-, \ E^-$ sources.
(The result obtained form $A_1^-, \ T_1^-, \ T_2^-$ sources instead  does not show
a significant difference.)
We observe that  (i) the central potential $V^{I=1}_{{\rm C};S=1}(r)$
is  repulsive,  (ii) the
tensor  potential $V^{I=1}_{\rm T}(r)$  is positive  and 
weak  compared to $V^{I=1}_{{\rm C};S=1}(r)$ and $V^{I=1}_{\rm LS}(r)$,
and  (iii)  the spin-orbit potential  $V^{I=1}_{\rm LS}(r)$
is  negative  and strong.   These  features agree qualitatively well with
those of the phenomenological potential in Ref.~\cite{Wiringa:1994wb}.

For both  spin-singlet and spin-triplet central  potentials, there may
be a  very weak  attractive pocket of  less than  a few MeV  at medium
distance $(r\simeq  1$ fm).  However, considering the  statistical and
systematic errors,  its existence should be carefully  examined in 
future studies.

We make a technical comment. We sometimes observe  large condition numbers for eq.~(\ref{eq:triplet}) (with three sources) near the spatial boundaries,  which gives rise  to points with  large statistical
errors at $r \simeq 1-1.5$ fm in Fig.~\ref{fig:VcVTVLS}. 

\subsection{Scattering phase shifts and effective potentials}
\label{sec.scat.phase}

\begin{figure}[!htb]
  \includegraphics[width=0.8\textwidth]{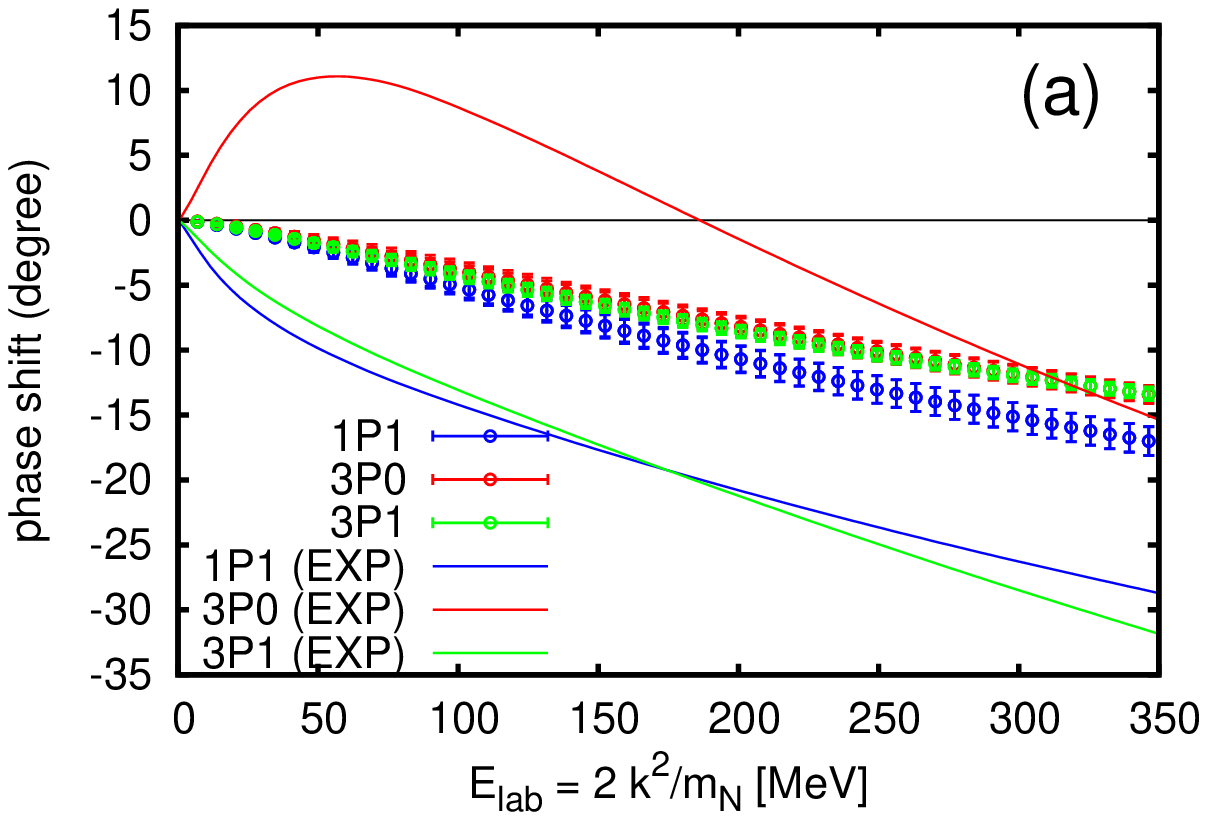}
\\
    \includegraphics[width=0.8\textwidth]{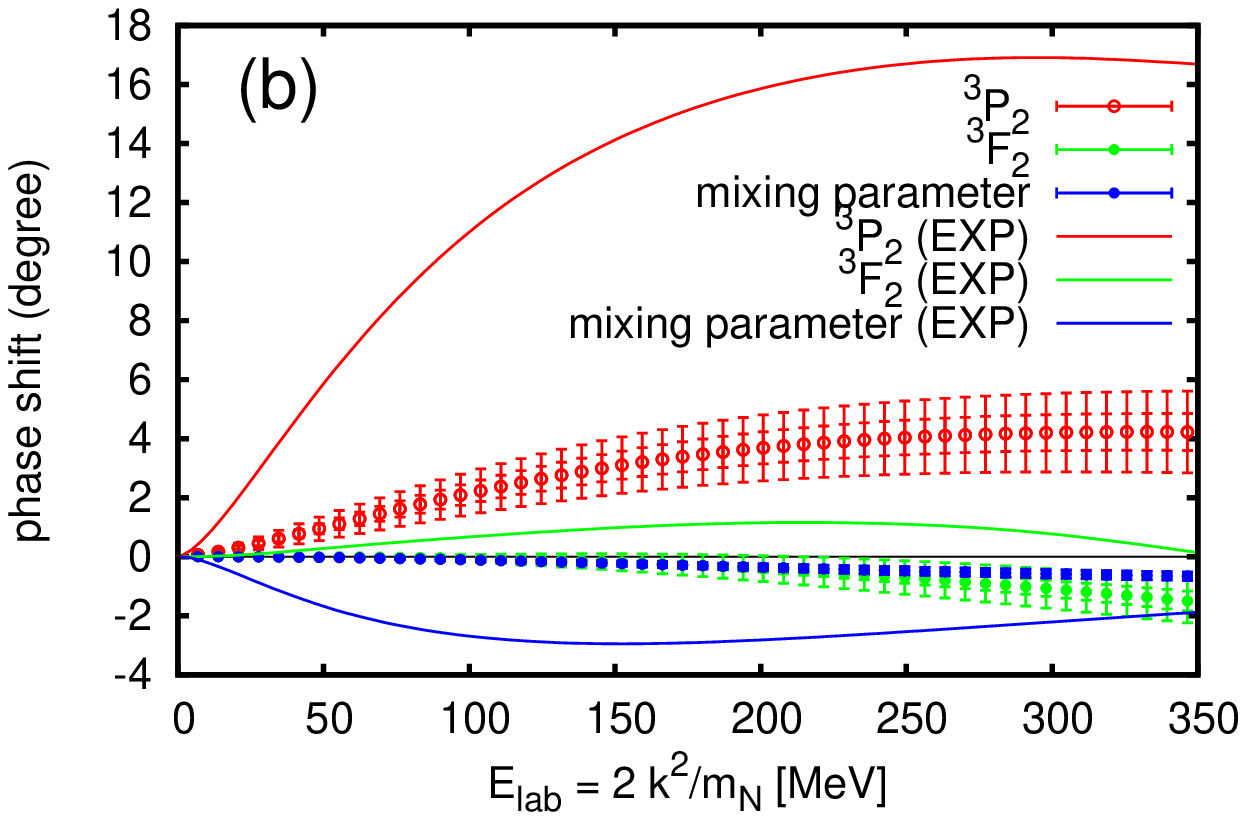}
  \caption{ The scattering phase shifts from Schr\"odinger equations by
    using the  potentials obtained at  $m_{\pi} \simeq 1133$  MeV from
    lattice QCD.
    (a) Phase shifts in $^1P_1$, $^3P_0$ and $^3P_1$ together with the
    experimental ones for comparison.
    (b) The phase shifts and  mixing parameter in the $^3P_2$--$^3F_2$
    channel together  with the experimental ones.  (Stapp's convention
    is adopted \cite{deSwart:1995ui}.)
    The inner error is statistical, while the outer one is statistical
    and systematic combined in quadrature.
%
}
 \label{fig:phase}
\end{figure}

\begin{figure}[!htb]
  \includegraphics[width=0.8\textwidth]{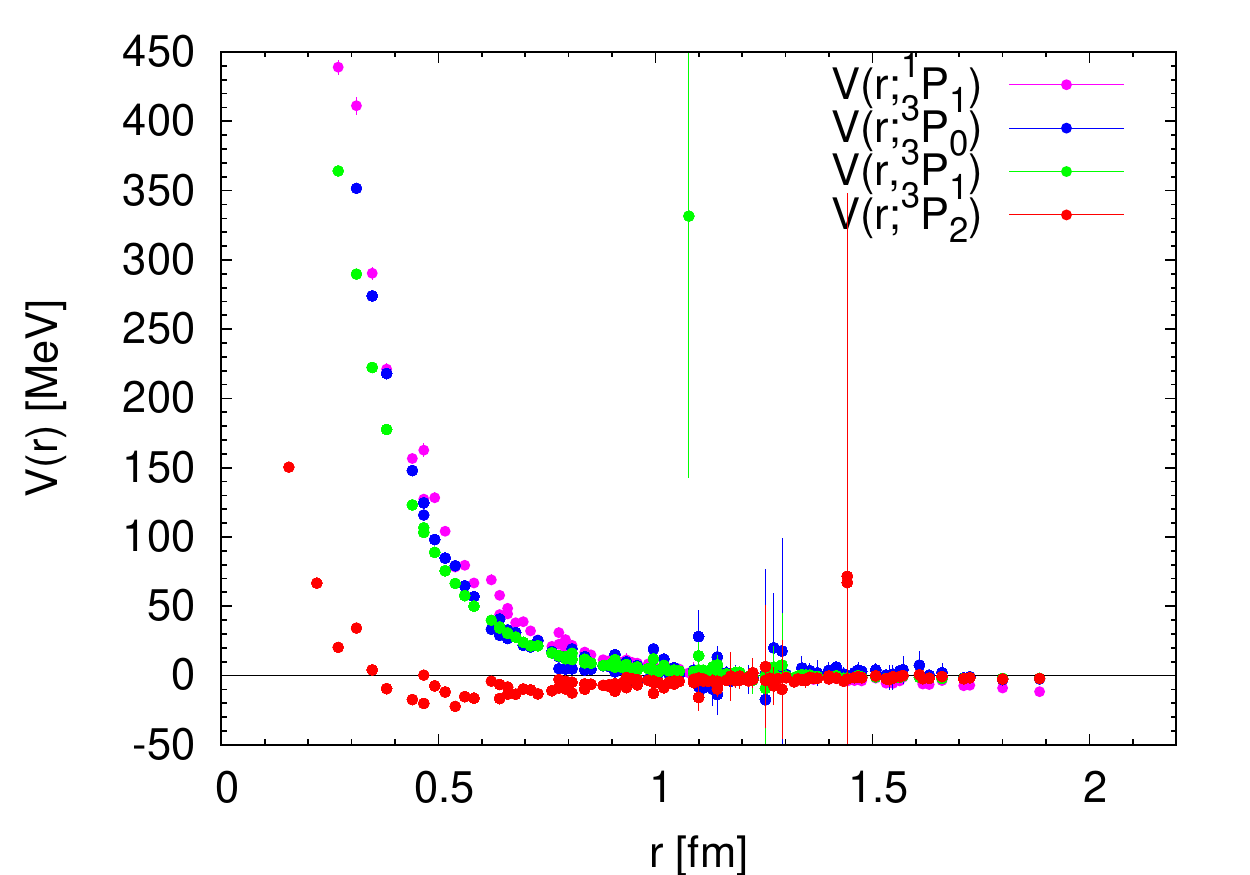}
  \caption{The potentials  for the $^1P_1$, $^3P_0$,  $^3P_1$ and  $^3P_2$
 channels   given in Eqs.~(\ref{eq:effective-potentials-1})-(\ref{eq:effective-potentials-3}) and below.}
 \label{fig:eff-pot}
\end{figure}

  For quantitative studies of the interactions, it is
  desirable to calculate not only the potential but also scattering
  phase shifts, since the potential is not a physical observable as
  mentioned above.  In this section, we therefore investigate a nature
  of interactions, by calculating scattering phase shifts from the
  obtained potentials.
  In particular, we study whether 
  the LS potential of Fig.~\ref{fig:VcVTVLS} leads to attractive
  behaviors in the scattering phase shifts
  in the $^3P_2$ channel
.

We calculate the scattering  phase shifts by solving the Schr\"odinger
equation with  the above potentials,  parameterized with multi-Gaussian
forms,  $v(r)   \equiv  \sum_{i=1}^{N_{\rm  gauss}}   a_i  \exp(-\nu_i
(r/b)^2)$ with $N_{\rm gauss}=3$ for the central and spin-orbit
potentials,  whereas   $v(r)  \equiv  a_1   (r/b)  \exp(-\nu_1
(r/b)^2) + a_2  (r/b)^3 \exp(-\nu_2 (r/b)^2)$ for
the tensor potential to mimic the short distance behavior, as shown in
\Fig{fig:VcVTVLS}.
Here,  a scaling  parameter  $b  \equiv 0.1555$  fm  is introduced  to
simplify the notation.
  The  uncorrelated fits  are performed  reasonably. The
  resultant   fit   parameters   and   $\chi^2$/dof   are   given   in
  Table~\ref{tab:fit}.

\begin{table}[tb]
\begin{center}
\begin{tabular}{|c|cccccc|c|}
\hline
channel & $a_1$ [MeV] & $a_2$ [MeV] & $a_3$ [MeV] & $\nu_1$ & $\nu_2$ & $\nu_3$ & $\chi^2$/dof \\
 \hline
 $V_{C;S=0}^{I=0} $&2173(268)&762(62)&236(65)  &11(2)     &2.1(0.3)    &0.6(0.1)  &1.7(0.8)\\
 $V_{C;S=1}^{I=1} $&421(122) &233(74)&397(16)  &11.5(0.4) &1.3(0.2) &3.9(0.1)     &1.1(1.0)\\
 $V_{T}^{I=1}    $&711(11)  &16(5)  &---       &2.6(0.2) &0.5(0.1) &---           &0.8(0.5)\\
 $V_{LS}^{I=1}   $&-45(17)  &-181(5)&-315(12)  &0.4(0.1) &1.4(0.2) &5.3(0.3)       &3.6(0.7)\\
 \hline
\end{tabular}
%
\caption{Fit parameters and $\chi^2$/dof.}
\label{tab:fit}
\end{center}
\end{table}

The  scattering  observables  are  obtained  from  the  long  distance
behaviors of linearly independent regular solutions,
and   are  shown   in   Fig.\ref{fig:phase}.   The   inner  error   is
statistical,  while  the  outer  one is  statistical  and  systematic
combined in quadrature.
Here,  to estimate  the systematic  error,  we take  into account  the
uncertainty arising  from the truncation of  the derivative expansion
and from the choice of fitting functions for the potentials.
To estimate  systematic errors associated  with the truncation  of the
derivative expansion, we calculate phase shifts also at $t-t_0=7$, and
take  differences  of  central  values  between  $t-t_0=8$  and  7  as
systematic errors.
  A dependence  of phase shifts  on a choice  of fitting
  functions for  the potentials is  estimated by changing  the fitting
  function to a Yukawa-type.
  It turns out  that the former dominates the  systematic error except
  that the latter dominates in the $^3F_2$ channel.
Although  the  magnitude  of  the   phase  shifts  obtained  from  our
potentials are smaller than the  experimental ones, general trends are
well reproduced except for the $^3P_0$  case at low energies.  The missing
attraction in the $^3P_0$  channel is likely due to the  weak tensor force
$V_{\rm T}$  caused by the  large pion  mass.  Among others,  the most
interesting feature  in Fig.\ref{fig:phase}  is the attraction  in the
$^3P_2$ channel, which  is directly related to  the paring correlation
of the neutrons inside the neutron stars.

     To obtain an intuitive understanding of the behavior of these phase shifts,
   we plot  the  potentials of the $^1P_1$, $^3P_0$, $^3P_1$ and
$^3P_2$ channels in Fig.\ref{fig:eff-pot},  as defined in  Eqs.~(\ref{eq:effective-potentials-1})-(\ref{eq:effective-potentials-3}) and below.
 Indeed, one can see that 
 $V(r;{^3P_2})$ has  a weak  repulsive core  surrounded  by an
 attractive well;  the attraction is driven by the strongly attractive 
  LS force, $V^{I=1}_{LS}(r)$ in  Fig.\ref{fig:VcVTVLS}.
 

    We give a comment  on the reliable energy region of
    these phase shifts. Through the time-dependent method, these phase
    shift are obtained  based on the NBS wave functions  at the energy
    points $E_{\rm lab}  \simeq 2 (2\pi/L)^2 \vec  n^2/m_N \simeq 230,
    460, 690, \cdots$ MeV with $\vec n \in \mathbb{Z}^3$.
%
    Except for these energy points and the point for $E_{\rm lab} = 0$
    where the  value of  the phase shift  vanishes by  definition, the
    phase  shifts  are  obtained   by  assuming  that  the  derivative
    expansion  is  converged  so  that  the  truncated  potentials  in
    \Eq{eq:OM-decomposition} do not depend on the energy.
    Because  the  contribution  from  $E_{\rm  lab}  \simeq  230$  MeV
    gradually dominates  the intermediate states in  ${\cal R}(\vec r,
    t;  {\mathscr J})$  as  $t$ increases,  the  most reliable  energy
    region of the phase shift is around $E_{\rm lab} \simeq 230$ MeV.
    Reliability for $E_{\rm lab} < 230$ MeV can be explicitly examined
    by enlarging  the spatial volume.  Reliability for  $E_{\rm lab} >
    230$ MeV  can be  examined by changing  relative weight  of excited
    states, which can be done either by studying the $t$ dependence of
    the  truncated potentials  or by  changing the  two-nucleon source
    operator.
    Note that  these arguments apply  to any choices  of interpolating
    fields.

\section{Conclusion}
We have made  a first attempt to determine NN potentials  up to NLO in
the  parity-odd sector,  which appears  in the $^1P_1$,  $^3P_0$, $^3P_1$,
$^3P_2$--$^3F_2$ channels.  Using $N_f=2$ CP-PACS gauge configurations
on  a $16^3  \times 32$  lattice ($a\simeq  0.16$ fm  and $m_\pi\simeq
1100$ MeV),  not only the central  and the tensor  potentials but also
the spin-orbit potential have been derived for the first time.
These potentials are constructed from NBS wave functions for $J^P=0^-,
1^-, 2^-$, which are generated  by using the momentum wall sources with
projections  based on the representation theory of the cubic group.

We have  observed that the  qualitative behavior of the resultant potentials
agree with those of phenomenological potentials:
For   the  spin-singlet sector,
  the   central   potential  
  $V^{I=0}_{{\rm C};S=0}(r)$  is repulsive  with a  strong repulsive  core at
short distance.
For   the  spin-triplet  sector, 
(i) the   central   potential  
$V^{I=1}_{{\rm C};S=1}(r)$ is also repulsive with a repulsive core at short
distance,
(ii) the  tensor  potential  $V^{I=1}_{\rm  T}(r)$ is  positive  and  quite
weak, and (iii) the  spin-orbit potential $V^{I=1}_{\rm LS}(r)$  is negative and
strong at short distance.

We  have then calculated scattering observables in the $^1P_1$, $^3P_0$, $^3P_1$
 and $^3P_2$--$^3F_2$ channels,
 by solving  Schr\"odinger  equations with these potentials.
It is interesting enough that  we obtain, from the first principle lattice QCD approach,  
attractive phase shift driven by the strongly attractive LS force in the $^3P_2$ channel, which has been known 
experimentally and has various implications in atomic nuclei and 
dense matter, 
 though the magnitude of the phase shift is still small due to the large quark mass in our calculation.
 This, together with the missing attraction in the $^3P_0$ channel,
  indicates an importance to carry out  simulations 
  at and around  the physical quark mass.

Numerical  calculations in this report  are  performed  on the  University  of  Tsukuba
Supercomputer-system   (T2K).    This  work   is   supported  in part by   the
JSPS Grant-in-Aid for Scientific Research (23540321, 24740144, 24740146), the JSPS Grant-in-Aid for Scientific Research on Innovative Areas(Nos.2004: 20105001, 20105003) and SPIRE (Strategic Program for Innovative Research).
We  are also  grateful for  authors  and  maintainer  of CPS++\cite{cps},  a
modified    version    of   which    is    used    for   this    work.
We  thank the CP-PACS  Collaboration and  ILDG/JLDG for  providing the  2-flavor QCD
gauge configurations\cite{hep-lat/0105015,jldg}.

\bibliographystyle{model1-num-names}


\end{document}